# АНАЛИЗ ИНФОРМАЦИИ О ПОЛЯХ ЗРЕНИЯ ЧЕЛОВЕКА С ИСПОЛЬЗОВАНИЕМ МЕТОДОВ МАШИННОГО ОБУЧЕНИЯ И ОЦЕНКА ИХ ТОЧНОСТИ

**Медведева Анастасия Игоревна**

*научный сотрудник учебно-научной лаборатории*

*искусственного интеллекта,*

*нейротехнологий и бизнес-аналитики*

*РЭУ им. Г. В. Плеханова,*

*Москва, Россия*

*E-mail: Medvedeva.AI@rea.ru*

**Бакуткин Валерий Васильевич**

*доктор медицинских наук, профессор,*

*генеральный директор ООО «МАКАО ИТ»,*

*Саратов, Россия*

*E-mail: bakutv@bk.ru*

Предмет исследования: изучение методов анализа периметрических снимков для диагностики и контроля заболеваний глаукомой.

Объект исследования: датасет, собранный на офтальмологическом периметре с результатами различных патологий пациентов, так как в офтальмологическом сообществе остро стоит вопрос контроля заболеваний и импортозамещения [5].

Цель исследования: рассмотреть различные методы машинного обучения, способные классифицировать глаукому. Это возможно благодаря классификатору, построенному после разметки датасета. Он способен определять по снимку, являются ли изображенные на нем поля зрения результатами воздействия на глаза глаукомы или же это другие зрительные заболевания. Ранее в работе [3] описывался датасет, который был собран на периметре «Tomey». Средний возраст обследованных пациентов составляет от 30 до 85 лет.

Методы исследования: методы машинного обучения для классификации результатов изображений (стохастический градиентный спуск, логистическая регрессия, случайный лес, наивный байес).

Основные результаты исследования: результатом исследования является компьютерное моделирование, способное определять по снимку, является ли результат глаукомой или иным заболеванием (бинарная классификация).

*Ключевые слова*: глаукома, машинное обучение, классификатор, поле зрения, периметрия.

# ANALYSIS OF HUMAN VISUAL FIELD INFORMATION USING MACHINE LEARNING METHODS AND ASSESSMENT OF THEIR ACCURACY

**Anastasia I. Medvedeva**

*researcher of the educational and scientific laboratory*

*of artificial intelligence,*

*neurotechnology and business analytics*

*Plekhanov Russian university of economics,*

*Moscow, Russia*

*E-mail: Medvedeva.AI@rea.ru*

**Valery V. Bakutkin**

*Doctor of Medical Sciences, Professor,*

*General Director LLC "MACAO IT"*

***Saratov**, Russia*

*E-mail: bakutv@bk.ru*

**This research was performed in the framework of the state task in the field of scientific activity of the Ministry of Science and Higher Education of the Russian Federation, project "Models, methods, and algorithms of artificial intelligence in the problems of economics for the analysis and style transfer of multidimensional datasets, time series forecasting, and recommendation systems design", grant no. FSSW-2023-0004.**

Subject of research: is the study of methods for analyzing perimetric images for the diagnosis and control of glaucoma diseases.

Objects of research: is a dataset collected on the ophthalmological perimeter with the results of various patient pathologies, since the ophthalmological community is acutely aware of the issue of disease control and import substitution. [5].

Purpose of research: is to consider various machine learning methods that can classify glaucoma. This is possible thanks to the classifier built after labeling the dataset. It is able to determine from the image whether the visual fields depicted on it are the results of the impact of glaucoma on the eyes or other visual diseases. Earlier in the work [3], a dataset was described that was collected on the Tomey perimeter. The average age of the examined patients ranged from 30 to 85 years.

Methods of research: machine learning methods for classifying image results (stochastic gradient descent, logistic regression, random forest, naive Bayes).

Main results of research: the result of the study is computer modeling that can determine from the image whether the result is glaucoma or another disease (binary classification).

## ВВЕДЕНИЕ

В офтальмологии является важным изучать поля зрения, так как их изменения могут указывать на наличие различных заболеваний, включая глаукому, миопию и другие неврологические расстройства [4]. С развитием технологий методы машинного обучения становятся все более актуальными для анализа и интерпретации этих данных. Говоря об искусственном интеллекте в медицине, важно учитывать и этические аспекты, чтобы не нарушать права пациентов [2].

В данной статье рассматривается применение методов машинного обучения для анализа полей зрения, оценка их точности и эффективности. Сосредоточимся на различных алгоритмах, таких как логистическая регрессия, случайный лес и других, а также проведем сравнительный анализ их производительности.

Целью исследования является не только выявление наиболее эффективных методов, но и понимание того, как эти технологии могут быть интегрированы в клиническую практику для улучшения диагностики и мониторинга заболеваний глаз.

В результате проведенного анализа важно заметить развитие методов диагностики и лечения заболеваний, связанных с изменениями в полях зрения, ведь потенциал машинного обучения особенно важен в области медицины. Повышая точность полученных данных, а также увеличивая количество испытуемых, стало возможным построение классификатора. Важно понимать, что всё это необходимо для улучшения жизнедеятельности пациентов и контроля течения заболевания.

## РЕЗУЛЬТАТЫ И ОБСУЖДЕНИЕ

Одной из задач машинного обучения, применимых к данному исследованию, является бинарная классификация. Алгоритмы анализируют входные данные и выдают предсказания о принадлежности объекта к одному из двух классов.

Этапы разработки классификатора:

1. Сбор данных

Как говорилось в работе [3], был собран и размечен датасет для определения координат точек на снимках. В данной статье каждый снимок был разделён на 2 глаза (каждый глаз отдельно). Собранный набор изображений глаз был помечен как «глаукома» и «иные заболевания» (рисунок 1);

**Рисунок 1.** Пример снимка правого глаза (OD) пациента, больного глаукомой

2. Предобработка изображений

Так как качество изображений было удовлетворительным (изображения были сняты на камеру с бумажного носителя или экрана монитора), они могли быть изменены для уменьшения шума, нормализации яркости и контрастности.

3. Обучение модели

Данные были разделены на обучающую и тестовую выборки. Обучение происходило с использованием извлеченных признаков. Обозначение 1 – глаукома, 0 – иные заболевания.

4. Прописывание гиперпараметров модели для повышения точности.

5. Рассмотрение и оценка точности модели.

Рассмотрим ключевые метрики оценки качества бинарной классификации, а именно precision, recall, F1-score и support. Эти показатели играют важную роль в анализе производительности классификаторов, особенно в задачах, где классы имеют неравномерное распределение.

Precision (точность) измеряет долю правильно предсказанных примеров глаукомы среди всех предсказанных положительных, что позволяет оценить, насколько классификатор надежен в своих положительных предсказаниях диагноза.

Recall (полнота) отражает долю правильно диагностированных снимков среди всех фактических положительных, что важно для понимания способности модели выявлять все положительные случаи.

F1-score представляет собой гармоническое среднее между precision и recall, обеспечивая сбалансированную оценку, особенно в ситуациях, когда необходимо учитывать как ложные срабатывания, так и пропуски.

Support указывает на количество фактических примеров каждого класса в тестовом наборе данных, что помогает интерпретировать другие метрики в контексте распределения классов [1].

В таблице 1 рассматривается применение метода логистической регрессии в рамках обучения модели на ранее описанном датасете. Определим взаимосвязь между независимыми переменными (признаками) и зависимой переменной (целевой переменной) с помощью логистической функции.

В результате этого accuracy (доля правильно классифицированных снимков (как с диагнозом «глаукома», так и иных заболеваний) к общему числу экземпляров) составляет 0,72, при этом support составляет 29. Случайный лес показал более низкий результат на то же количество support, а именно 0,69. Стохастический градиентный спуск и случайный лес составляют 0,69 (см. таблицы 3, 4, а самый худший результат показал наивный байес – 0,45 (см. таблицу 2), хотя преимущество наивного байеса заключается в его простоте и скорости обучения, что делает его эффективным для больших наборов данных. Его предположение о независимости признаков может ограничивать точность в нашей задаче.

**Таблица 1.** Обучение модели методом логистической регрессии

| Признак | Precision | Recall | F1-score | Support |
|---------|-----------|--------|----------|---------|
| 0 | 0,69 | 0,69 | 0,69 | 13 |
| 1 | 0,75 | 0,75 | 0,75 | 16 |
| Accuracy | 0,72 | | | 29 |

**Таблица 2.** Обучение модели – наивный байес

| Признак | Precision | Recall | F1-score | Support |
|---------|-----------|--------|----------|---------|
| 0 | 0,20 | 0,08 | 0,11 | 13 |
| 1 | 0,50 | 0,75 | 0,60 | 16 |
| Accuracy | 0,45 | | | 29 |

**Таблица 3.** Обучение модели – случайный лес

| Признак | Precision | Recall | F1-score | Support |
|---------|-----------|--------|----------|---------|
| 0 | 0,61 | 0,85 | 0,71 | 13 |
| 1 | 0,82 | 0,56 | 0,67 | 16 |

| Accuracy | 0,69 | | 29 |
|----------|------|---|-----|

| Признак | Precision | Recall | F1-score | Support |
|---------|-----------|--------|----------|---------|
| 0 | 0,67 | 0,62 | 0,64 | 13 |
| 1 | 0,71 | 0,75 | 0,73 | 16 |
| Accuracy | 0,69 | | | 29 |

## ЗАКЛЮЧЕНИЕ И ВЫВОДЫ

В результате данного исследования было рассмотрено несколько методов обучения модели. Лучше всего себя показала логистическая регрессия с точностью 0,72, из чего можно сделать вывод, что для задачи классификации снимков этот метод наиболее предпочтителен. Логистическая регрессия и бинарная классификация являются важными инструментами для решения задач по определению диагнозов в нашей работе.

Это простая модель, которая легко интерпретируется. Она позволяет понять, как изменения в независимых переменных влияют на вероятность наступления события, также это идеально подходит для задач бинарной классификации, где необходимо предсказать одно из двух возможных состояний (глаукома или иные заболевания).

В дальнейшем планируется построить матрицы ошибок и ROC-кривые AUC для детального рассмотрения качества классификации. Несомненно, возможно исследовать обучение данного датасета на сверточных нейронных сетях, а также его расширение и изучение группы паталогически здоровых пациентов.

# СПИСОК ЛИТЕРАТУРЫ